\documentclass[]{aa}

\usepackage{graphicx}
\usepackage{color}

\begin{document}

\title{A dynamo mechanism as the potential origin of the long cycle in Double Periodic Variables}

\titlerunning{The dynamo cycle in DPVs}
\authorrunning{Schleicher \& Mennickent}

\author
  {Dominik R.G. Schleicher
  \inst{1}
  \and
  Ronald E. Mennickent
  \inst{1}
  }

\institute{Departamento de Astronom\'ia, Facultad Ciencias F\'isicas y Matem\'aticas, Universidad de Concepci\'on, Av. Esteban Iturra s/n Barrio Universitario, Casilla 160-C, Concepci\'on, Chile; 
\email{dschleicher@astro-udec.cl}
}

\date{\today}

\abstract{
The class of Double Period Variables (DPVs) consists of close interacting binaries, with a characteristic long period {that is an order of magnitude longer than the corresponding orbital period, many of them with a characteristic ratio of about $3.5\times10^1$.} We consider here the possibility that the accretion flow is modulated as a result of a magnetic dynamo cycle. Due to the short binary separations, we expect the rotation of the donor star to be synchronized with the rotation of the binary due to tidal locking. {We here present a model to estimate the dynamo number and the resulting relation between the activity cycle length and the orbital period, as well as an estimate for the modulation of the mass transfer rate. The latter is based on Applegate's scenario, implying cyclic changes in the radius of the donor star and thus in the mass transfer rate as a result of magnetic activity. Our model is applied to a sample of {17} systems with known physical parameters, 10 also with known orbital periods. In spite of the uncertainties of our simplified framework, the results show a reasonable agreement, indicating that a dynamo interpretation is potentially feasible. At the same time, we note that the orbital period variations resulting from Applegate's model are sufficiently small to be consistent with the data. We conclude that both larger samples with known physical parameters as well as potential direct probes of the magnetism of the donor star, including cold spots as well as polarization, will be valuable to further constrain the nature of these systems.}
}

\maketitle

\section{Introduction}
The study of interacting binary systems is particularly important as a probe of stellar physics and to understand the evolution of close binaries. An important class of these systems includes the Algol-type variables, which are semi-detached binary systems of intermediate stellar masses. {As originally explained by \citet{Crawford55} and later confirmed via evolutionary calculations by \citet{Kippenhahn67}}, the mass ratio distribution between the evolved star and the main sequence star indicates that severe interactions must have taken place in the binary system to account for the higher mass of the main sequence star, including significant amounts of mass transfer \citep[see also][]{Sarna93, Rensbergen11, Mink14}. After fast mass exchange as a result of Roche lobe overflow, the lobe-filling donor star is thus significantly less massive than the primary \citep{Eggleton06}.

\subsection{{Long cycles in interacting semi-detached binaries}}

{
A particularly interesting feature in the interacting semi-detached binaries is the observation of long cycles. The presence of such cycles is known since a long time, as \citet{Gaposchkin44} inferred the presence of a long cycle in RX~Cas, \citet{Lorenzi80b, Lorenzi80a} demonstrated the presence of a long cycle in AU~Mon, and \citet{Guinan89} inferred it for $\beta$~Lyr. Nowadays, much larger samples of binaries with long cycles are known, and the above mentioned systems as well as other binaries with well-known physical parameters were for instance recently revisited by \citet{Harmanec15}.}

{The interpretation of these long cycles however remains enigmatic until this point. For instance,} \citet{Peters94}, associated the long-period variation of AU\,Mon to changes in the mass transfer rate due to cyclic pulsations of the donor star, but no compeling reason for this possible oscillation is provided. In this paper, our main concern are the Double Periodic Variables (DPVs), a sub-class of the Algols consisting of particularly massive stars, where the primary has a typical mass of at least $\sim7$~M$_\odot${, and the secondary is a star filling its Roche lobe}. This class of binary stars was initially discovered based on the inspection of large scale photometric surveys of emission-line objects in the Magellanic Clouds. \citet{Mennickent03} published a first list with initially 30 objects exhibiting roughly sinusoidal periodic light variations with periods from 140 to 960~days. They further showed periodic changes with periods from 2.4 to 15.9~days, including sinusoidal, ellipsoidal and eclipsing light curves. They found a characteristic relation between the long and short period of about $P_{\rm long}=35.17\times P_{\rm short}$. This relation and the existince of the two periods has given rise to the name of Double-Periodic Variables.

The first spectroscopic data for a sample of DPVs in the Magellanic Clouds has been obtained by \citet{Mennickent05}, finding that the optical spectrum was dominated by Balmer and helium absorption lines and a continuum with a blue or sometimes flat slope. One object in the sample showed a characteristic shortening of the long cycle by about $20\%$ in some of the cycles, suggesting that the periodicity is not strict. Similarly, \citet{Mennickent06} examined OGLE and MACHO light curves of DPVs spanning 11 years, finding excursions from strictly periodic variability and a number of cases with either increasing or decreasing cycle lengths. Since then, additional DPV systems have been found both in the Galaxy \citep{Mennickent12b}, in the Magellanic Clouds \citep{Poleski10, Pawlak13} {as well as towards the Galactic bulge \citep{Soszynski16}}. Interestingly, deeper primary eclipses are observed during the long cycle and in one system the secondary eclipse disappears at some epochs \citep{Poleski10}.

The DPV OGLE05155332-6925581 is a semi-detached system with H$\alpha$ emission and a luminous accretion disk similar to $\beta$~Lyr; a preliminary interpretation was given for their long cycle in terms of cycles of mass loss in the system, probably feeding a circumbinary disc \citep{Mennickent08}, a kind of decretion disk as described by \citet{Tutukov04}. A similar explanation was put forward by \citet{Desmet10} in the context of AU~Mon, challenging the \citet{Peters94} interpretation; CoRoT space photometry shows that the orbital light curve remains practically constant through the long cycle. Other explanations that were previously considered included the potential influence of disk winds, as detected for instance in V~393~Sco \citep{Mennickent12} and HD~170582 \citep{Mennickent15}. Similarly, the precession of the circumprimary disk could in principle lead to such effects via a 3:1 resonance \citep{Mennickent03}, but the observed double emission lines show no such indications. The relation to the cool donor star became however more obvious from the four-colour light curves published by \citet{Michalska10} for several DPVs, showing that in all cases the amplitude of the long-term cyclic changes were growing towards longer wavelengths.

In a recent study, \citet{Mennickent16} have attempted to further clarify the physical origin of the DPV phenomenon as well as the nature of their disks, which were frequently inferred around the more massive stars. As it is well-known from the studies by \citet{Kriz70, Kriz71, Kriz72}, there is an important distinction between direct impact in shorter period systems, and the formation of an accretion disk if the impact parameter is initially larger than the radius of the primary. \citet{Lubow75} have quantified this criterion in terms of the Lubow-Shu critical radius. If the radius of the primary is smaller than this critical radius, it means that an accretion disk will form around the star, while the absence of a disk is expected if the stellar radius is larger than the critical radius, and the accretion flow will then directly hit the stellar surface. As found by \citet{Mennickent16}, the critical radius appears to be very similar to the radii of the primaries for the known DPV systems, suggesting that the accretion flow hits the survace of the primary in an almost tangential manner. 

The latter allows in particular to distinguish the DPVs from the class of W Serpentis stars which include similar stellar masses and orbital periods, but where the radius of the primary is typically much smaller than the critical radius. This class of stars consists of a hot primary surrounded by an optically thick accretion disk \citep{Plavec80a, Young82}, and is characterized by strong ultraviolet emission lines of highly excited species, including He~II, C~II, Al~III, Fe~III, C~IV, Si~IV and N~V, potentially formed in a super corona produced via mass transfer and accretion \citep{Plavec82, Plavec89}. They show significant variations of the orbital period and appear to be strongly interacting. 

In the following, we aim to clarify the overall picture of the DPVs that has emerged over time. They consist of a B-type star plus a less massive  ($1-3$~M$_\odot$) companion filling the Roche lobe, leading to mass overflow to the more massive star. The more massive star is surrounded by a massive and optically thick disk, where the disk radius is comparable to the radius of the primary star. Based {on} calculations following the evolutionary tracks of these systems \citep{Mennickent16}, it appears very likely that the primary has accreted a significant mass from the donor, thereby gaining a large amount of angular momentum. Even if not expected within the Lubov-Shu picture, the formation of an accretion disk is thus plausible, as likely not all of the angular momentum can be accreted onto the primary star, and a mechanism like a disk wind may thus be beneficial to loose some of the angular momentum. The eccentricity in DPVs is generally compatible with zero; in very few cases it is very small and consistent with the effects of the mass streams \citep{Lucy05}. The latter is expected as a result of rapid circularization via dynamical tides \citep{Tassoul87, Zahn89a, Zahn89b}.

\subsection{{Evidence for magnetic activity cycles in various types of binaries}}

{In this paper, we consider magnetic activity cycles as a potential origin of the long period in DPVs. These long cycles have been inferred via changes in the light curves, rather than orbital period variations, and we will in fact show that the expected orbital period variations are negligible for the DPVs.}  {In the following, we provide a brief overview on what is known regarding magnetic activity in different types of binary systems. In fact, already the Algol system itself shows a characteristic 32 year cycle originally reported by \citet{Soderhjelm80}, and its connection to magnetic activity has been confirmed via radio observations by \citet{Peterson10}. The possibility of such magnetic activity cycles has been put forward by  \citet{Sarna97} for the more general class of Algol-type systems, and was more recently revisisted by \citet{Soker02}. }

{In the sub-class of RS CVn stars, which are detached binaries typically composed of a chromospherically active G or K star, the orbital period variations were explored by \citet{Lanza98} and \citet{Lanza99} and related to the presence of a magnetic activity cycle. The latter can be  explained via Applegate's mechanism, explaining the orbital period variations as a result of quasi-periodic changes of the quadrupole moment due to the internal redistribution of angular momentum via magnetic fields \citep[][see \citet{Volschow16} for an updated version of the formalism]{Applegate87}. As further described by \citet{Bolton89} and \citet{Meintjes04}, the presence of a dynamo may modulate the mass transfer rate in Algol systems, leading to a characteristic impact of the dynamo cycle on the luminosity of the resulting hot spot (even if there is not a hot spot, but rather a hot belt, as proposed by \citet{Bisikalo10}, the resulting behavior will be similar). }

{Orbital period modulations have also been observed for a significant number of post-common-envelope binaries (PCEBs), including V471~Tau, DP~Leo and QS~Vir \citep{Zorotovic13}. While there is still an ongoing debate if also planets might be present in these systems, at least for some of them the presence of magnetic activity is clearly confirmed. For V471~Tau, magnetic activity was shown via photometric variability, flaring events and H$\alpha$ emission along with a strong X-ray signal \citep{Kaminski07, Pandey08}. For DP~Leo, the presence of magnetic activity is indicated via X-ray observations \citep{Schwope02}, and in case of QS~Vir, it has been inferred via detections of Ca~II emission and Doppler Imaging \citep{Ribeiro10} as well as observed coronal emission \citep{Matranga12}. A theoretical model for dynamos in these type of systems was put forward by \citet{Ruediger02}.
}

{Even for single main-sequence stars, characteristic relations between the stellar rotation period and the dynamo cycle are known, and have been reported by \citet[e.g.][]{Saar99, Boehm07}. These results are frequently interpreted in terms of dynamo models, as already \citet{Soon93} and \citet{Baliunas96} suggested a correlation between rotation velocity, activity period and dynamo number $D=\alpha \Delta \Omega d^3/\eta^2$. Here $\alpha$ is a measure of helicity, $\Delta\Omega$ the large-scale differential rotation, $d$ the characteristic length scale of convection and $\eta$ the turbulent magnetic diffusivity in the star. Beyond the main sequence, successful dynamo simulations have been pursued for AGB stars by \citet{Blackman01} and \citet{Dorch04}, while polarized water maser observations have confirmed the presence of $\sim10-100$~G magnetic fields around evolved isolated stars \citep{Vlemmings02, Vlemmings05, Leal13}. In case of the asymptotic giant branch star W43A, \citet{Vlemmings06} further detected the presence of a collimated jet as a result of magnetic activity. }

Our paper is structured as follows: In section~\ref{dynamo}, we present a dynamo model for DPVs including the formalism to estimate the dynamo number. Section~\ref{Applegate} presents a mechanism based on Applegate's scenario to explain a time-dependent mass transfer as a result of the dynamo cycle. In section~\ref{comp}, we compare our models to a sample of currently discussed DPV candidates. A final discussion is given in section~\ref{discussion}.

\section{A dynamo model for DPVs}\label{dynamo}
{
In the context of dynamo models, the dynamo cycle $P_{\rm cycle}$ is related to the rotation period $P_{\rm rot}$ via a relation of the form \citep{Soon93, Baliunas96}
\begin{equation}
P_{\rm cycle}=D^\alpha P_{\rm rot},\label{cycle}
\end{equation}
with $D$ the dynamo number and $\alpha$ a power-law index, with typical values of $\alpha$ between $\sim\frac{1}{3}$ and $\sim\frac{5}{6}$. Observational studies of single main-sequence stars have pointed towards an index $\alpha\sim0.25$ for two parallel-branches \citep{Saar99}, i.e. implying two different normalizations, while recent simulations by \citet{Dube13}, along with data for BY~Dra and W~UMa, hint at the possible existence of a super-active branch with $\alpha\sim-0.2$. The precise value of the power-law index is thus still controversial, and may also depend on the astrophysical object under consideration. In the following, we will thus aim to explore the dependence of the dynamo number $D$ on the stellar parameters, while the power-law index $\alpha$ will be determined by comparison with a sample of available objects. 

For this purpose, we express the dynamo number $D$ in terms of the Rossby number Ro as $D=$Ro$^{-2}$. Followig \citet{Soker00}, the latter is given as\begin{equation}
\mathrm{Ro}=9\left( \frac{v_c}{10\,{\mathrm{km/s}}} \right)\left( \frac{H_p}{40\,R_\odot} \right)^{-1}\left( \frac{\omega}{0.1\omega_{\rm Kep}} \right)^{-1}\left( \frac{P_{\rm Kep}}{\rm yr} \right),\label{Rossby}
\end{equation}
with $v_c$ the convective velocity, $H_p$ the pressure scale height, $\omega$ the angular velocity, $\omega_{\rm Kep}$ the Keplerian angular velocity and $P_{\rm Kep}$ the Keplerian orbital period of a test particle on the surface of the donor star. Assuming tidal locking \citep{Tassoul87, Zahn89a, Zahn89b}, the angular velocity of the donor star is equal to the angular velocity of the binary. We thus have\begin{equation}
\omega=\sqrt{\frac{G(M_1+M_2)}{a^3}},
\end{equation}
where $a$ denotes the separation of the binary, and we assume a circular orbit. As the radius of the donor stars fills the Roche lobe, it can be evaluated as \citep{Paczynski71} 
\begin{equation}
R_{\rm Roche}=0.46224a\left( \frac{M_2}{M_1+M_2} \right)^{1/3}\propto a \left( \frac{1}{1+q^{-1}} \right)^{1/3}\label{Roche2}
\end{equation}
for $q=M_2/M_1<0.8$. We assume that the scale height $H_p$ can be described as a fraction $\epsilon_H$ of the Roche radius $R_{\rm Roche}$. The Keplerian angular velocity on the surface of the donor star then follows as\begin{equation}
\omega_{\rm Kep}=\sqrt{GM_2/R_2^3}\sim3.2\omega,\label{omega}
\end{equation}
implying $\omega\sim0.31\omega_{\rm Kep}$. The Kepler period of a test particle at the surface of the donor star is\begin{equation}
P_{\rm Kep}=\frac{2\pi}{\omega_{\rm Kep}}=0.31\times\frac{2\pi}{\omega}=0.31\times2\pi\sqrt{\frac{a^3}{G(M_1+M_2)}}.\label{Kepler}
\end{equation}
A central ingredient to evaluate the Rossby number is then the convective velocity $v_c$. Under conditions of convective energy transport, it can be shown  that \citep{Kippenhahn12} \begin{equation}
v_c=v_s\sqrt{\nabla-\nabla_{ad}},\label{vc}
\end{equation}
with $v_s$ the speed of sound in the stellar interior, $\nabla=\frac{d\ln T}{d\ln r}$ the physical temperature gradient and $\nabla_{ad}=\left(\frac{d\ln T}{d\ln r}\right)_{ad}$ the temperature gradient under adiabatic conditions. From mixing length theory, we have\begin{equation}
F_{\rm conv}=\rho c_P T\left( \frac{l_m}{H_p} \right)^2\sqrt{\frac{1}{2}gH_p}(\nabla-\nabla_{ad})^{3/2},\label{Fconv}
\end{equation}
with $\rho$ the density, $F_{\rm conv}$ the convective energy flux, $C_P$ the heat capacity at constant pressure, $T$ the temperature, $l_m$ the mixing length and $g$ the gravitational acceleration. In the following, we will estimate all quantities in Eq.~(\ref{Fconv}) with the representative quantities inside the star. We have\begin{eqnarray}
F_{\rm conv}&\sim& \frac{L_2}{4\pi R_2^2},\quad  \rho\sim\frac{3M_2}{4\pi R_2^3}, \quad T\sim\frac{\mu}{\mathcal{R}}\frac{GM_2}{R_2},\\
C_P&\sim&\frac{5}{2}\frac{\mathcal{R}}{\mu}, \quad \sqrt{gH_p}=\sqrt{\frac{\mathcal{R}}{\mu}T}\sim\sqrt{\frac{GM_2}{R_2}},\label{two}\\
 g&\sim&\frac{GM_2}{R_2^2},\label{three}
\end{eqnarray}
with $L_2$ the luminosity of the donor star, $\mathcal{R}$ the gas constant and $\mu$ the mean molecular weight. Inserting these expressions in Eq.~(\ref{Fconv}) and solving for $(\nabla-\nabla_{ad})$, we obtain\begin{equation}
\nabla-\nabla_{ad}=\left( \frac{2\sqrt{2}}{15} \right)^{2/3}\frac{L_2^{2/3}R_2^{5/3}}{GM_2^{5/3}}\left( \frac{l_m}{H_P} \right)^{-4/3}.\label{nabla}
\end{equation}
{Combining Eq.~\ref{two} and \ref{three}, it is straightforward to show that $H_P\sim R_2$. While this may be a rather crude approximation, it results from the assumption of considering only average properties within the cool companion.}
The sound speed in the interior can be evaluated as\begin{equation}
v_s=\sqrt{\frac{GM_2}{R_{\rm 2}}}.\label{vs}
\end{equation}
The mixing length is assumed to be of the order of the pressure scale height $H_P$. We here parametrize it as $H_P=\epsilon_H R_2${, though we will for simplicity assume $\epsilon_H=1$}. The Rossby number can thus be evaluated as\begin{equation}
{\rm Ro}=11.5\frac{v_cP_{\rm Kep}}{\epsilon_H R_{\rm 2}}\frac{R_\odot}{{\rm km/s\ yr}}.
\end{equation}
Inserting this in Eq.~(\ref{cycle}), we obtain
\begin{equation}
P_{\rm cycle}=\left( 11.5\frac{v_cP_{\rm Kep}}{\epsilon_H R_{\rm 2}}\frac{R_\odot}{{\rm km/s\ yr}} \right)^{-2\alpha} P_{\rm rot}.\label{Pcycle}
\end{equation}
{Inserting Eqs.~(\ref{nabla}) and (\ref{vs}) into Eq.~(\ref{Pcycle}), we obtain}
\begin{eqnarray}
P_{\rm cycle}&=&P_{\rm rot}\left( 11.5 \left( \frac{2\sqrt{2}}{15} \right)^{1/3} \frac{R_\odot}{\rm yr}  \right)^{-2\alpha}\label{Pfin} \\
&\times&  \left(\frac{L_2^{2/3}R_2^{2/3}}{M_2^{2/3}}\left( \frac{l_m}{H_P} \right)^{-4/3}  \left( \frac{ P_{\rm Kep}}{\epsilon_H R_{\rm 2}} \right)^2 \right)^{-\alpha}.\nonumber
\end{eqnarray}
In the following section, we will describe the potential implicactions for the mass transfer as a result of Applegate's model, and subsequently pursue a systematic comparison of our model with the  available data.
}

\section{Mass transfer as a result of Applegate's scenario}\label{Applegate}
{ As it is well-known from Applegate's model, the presence of a dynamo in the stellar interior can have a relevant impact on the internal structure, affecting the angular momentum distribution and the stellar quadrupole moment \citep{Applegate87}. As shown already in the previous section, the systems considered here are very likely to exhibit such a dynamo mechanism, both due to their rapid rotation as a result of tidal locking, as well as the high convective velocities in the giant star phase. As the donor stars also fill the Roche lobe, we expect that the mass transfer rate in the system will be particularly susceptible even to small changes in the stellar structure, thus implying cyclic changes on the timescale of the stellar dynamo. 

In the following, we adopt the finite shell framework for Applegate's model, considering the angular momentum exchange between a finite outer shell and the inner part of the star, as originally derived by \citet{Brinkworth06} and subsequently extended by \citet{Volschow16}. In this framework, the relative period change $\Delta P/P_{\rm orb}$, with $P_{\rm orb}$ the period of the binary, is given as\begin{eqnarray}
\frac{\Delta P}{P_{\rm orb}}&=&0.91\times10^{-7}\left(\frac{\Delta E}{E_{\rm sec}}\right)\left(\frac{a}{R_\odot}\right)^{-2}\left(\frac{M_{\rm 2}}{M_\odot} \right)^{-2}\nonumber\\
&\times&\left(\frac{R_{\rm 2}}{R_\odot} \right)^{3}\left( \frac{P_{\rm cycle}}{yr}\right)\left( \frac{L_{\rm 2}}{L_\odot} \right).\label{dpP}
\end{eqnarray}
Here $\Delta E$ denotes the required energy for the structural changes inside the star within one dynamo period $P_{\rm cycle}$, $E_{\rm sec}$ is the energy produced inside the star during that time, $R_2=R_{\rm Roche}$ the radius of the secondary and $L_2$ its luminosity, which can be estimated from the effective temperature $T_{\rm eff}$ and the Stefan-Boltzmann constant $\sigma_{SB}$ as\begin{equation}
L_2=4\pi \sigma_{SB} R_{\rm Roche}^2T_{\rm eff}^4.
\end{equation}
From Eq.~{\ref{dpP}}, we already see that the strength of the Applegate effect is limited by the amount of energy that is available for structural changes within the star, with a maximum ratio of $\Delta E/E_{\rm sec}=1$. The latter will be adopted in the following to derive an upper limit on the potential effects. These results can be linearly rescaled to account for other possible ratios.

To assess the implications for the stellar interior, we adopt the formula \citep{Applegate92, Volschow16}\begin{equation}
\frac{\Delta P}{P_{\rm orb}}=-\frac{9\Delta Q}{a^2 M_2},
\end{equation}
with $\Delta Q$ the change in the quadrupole moment that is driving the period variation. Assuming that the initial quadrupole moment of the donor is approximately $Q\sim M_2 R_aR_b$, with $R_a$ and $R_b$ the extension of the donor along two axis and at least approximately $R_a\sim R_b\sim R_{\rm Roche}$, a change by $\Delta R$ along one axis will lead to a change \begin{equation}
\Delta Q = M_2 R_{\rm Roche}\Delta R,
\end{equation}
implying that \begin{equation}
\Delta R=\frac{\Delta Q}{M_2R_{\rm Roche}}.
\end{equation}
To estimate the mass that is transported to the primary as a result of the change $\Delta R$, we calculate the effective cross section of the flow as \citep{Ritter88}\begin{equation}
A_{\rm eff}=\frac{2\pi\mathcal{R}T_{\rm eff}R_2^3}{GM_2\mu}F(q^{-1}),
\end{equation}
with $\mathcal{R}$ the gas constant and $\mu$ the mean molecular weight, for which we adopt a value of 2 assuming predominantly ionized gas. As stated by \citet{Ritter88}, the function $F$ can be evaluated as\begin{equation}
F(q^{-1})=1.23+0.5\log(q^{-1})
\end{equation}
for $0.5\leq q^{-1}\leq 10$. We derive an upper limit on the potential mass transferred to the primary by considering the volume $A_{\rm eff}\Delta R$, multiplied with the density $\rho_{\rm Roche}$ at $R_2\sim R_{\rm Roche}$. We assume here a typical envelope structure similar to AGB stars, where a small mass $M_{\rm core}$ forms the central core, while most of the stellar mass is within the envelope $M_{\rm env}\sim M_2$. A typical power-law density profile is then given as \citep{Soker92, Kashi11, Schleicher14} \begin{equation}
\rho(r)=Ar^{-2},
\end{equation}
with $A=(M_2-M_{\rm core})/(4\pi R_{\rm Roche})$. The mean density in the envelope is then given as\begin{equation}
\bar{\rho}=\frac{M_2-M_{\rm core}}{\frac{4\pi}{3}R_{\rm Roche}^3},
\end{equation}
while the density at $R_{\rm Roche}$ follows from\begin{equation}
\rho(R_{\rm Roche})=AR_{\rm Roche}^{-2}=\frac{M_2-M_{\rm core}}{4\pi R_{\rm Roche}^3}=\frac{1}{3}\bar{\rho}.
\end{equation}
As a result, the mass transported through the effective area due to the change $\Delta R$ is given as
\begin{equation}
\Delta M=A_{\rm eff}\Delta R\frac{1}{3}\bar{\rho}\sim\frac{1}{3}A_{\rm eff}\Delta R\frac{M_2}{\frac{4\pi}{3}R_{\rm Roche}^3},
\end{equation}
where in the last step we assumed that the core mass will be negligible. The average mass transfer rate during one activity period $P_{\rm cycle}$  is then\begin{equation}
\dot{M}=\frac{\Delta M}{P_{\rm cycle}}.
\end{equation}
We note that this presents an estimate for the maximum averaged magnitude of the mass transfer rate based on cyclic variations of the stellar quadrupole moment as predicted in Applegate's model, specifically employing the framework outlined by \citet{Volschow16}. This mechanism thus corresponds to a modulation of the mass transfer rate and can give rise to a cyclic variation due to the accretion luminosity, as\begin{equation}
L_{\rm acc}=\frac{GM_1\dot{M}}{R_1},
\end{equation}
with $R_1$ the radius of the primary star and $\dot{M}$ the accretion rate. We note that such a relation holds rather independently of whether the accretion luminosity is released via a hot spot or a hot belt, as proposed by \citet{Bisikalo10}, as the released energy is determined by the released energy from the gravitational potential. The variation occuring on the cycle period thus needs to be discriminated from other variations occuring due to binary evolution, typically occuring on timescales of a few $100$~years in the regime of strong interactions. We also note that the values given here correspond to averages, therefore the actual variation at a given time can be smaller or larger.
}

\begin{table*}[htbp]
\begin{center}
\caption{Physical parameters of the systems in our sample, including primary and secondary masses $M_1$ and $M_2$, the mass ratio $q=M_2/M_1$, the secondary radius $R_2$, the effective temperature of the secondary $T_{\rm 2}$, the orbital period $P_{\rm orb}$ as well as the available information on the presence of a cycle period. We note that radii marked with $^*$ have been calculated using the Roche lobe formula (Eq.~\ref{Roche2}). Information regarding the presence and duration of a cycle period are based on \citet{Harmanec15} and \citet{Mennickent16}. {Information for the highly active system U~Cep are based on \citet{Manzoori08}. The symbol $^{**}$ refers to the full name OGLE 05155332-6925581 of iDPV.} }
\begin{tabular}{c|c|c|c|c|c|c|c}\label{DPVsystems}
Binary & $M_1$ & $M_2$  & q & $R_{2}$ & $T_{\rm 2}$ & $P_{\rm orb}$ & cycle\\ 
 & [$M_\odot$] & [$M_\odot$] & &  $[R_\odot]$ & [K] & [d] & [d] \\ \hline
U~Cep & 4.938 & 3.22 & 0.652 & 7.05 & 3535 & 3.38 & 515-26663\\
UX~Mon & 3.38 & 3.9 & 1.15 & 9.95$^*$ & 5990 & 5.90 & no\footnotemark \\
DQ~Vel  & 7.3 & 2.2 & 0.31  & 8.4 & 9350 & 6.08 & 189\\
V448~Cyg & 24.7 & 13.7 & 0.55 & 16.17$^*$ & 20340 & 6.52 & no\\ 
CX~Dra & 7.3 & 1.7 & 0.23 & 13.35$^*$ & 6500 & 6.70 & yes\\
TT~Hya & 2.77 & 0.63 & 0.23 & 4.3$^*$ & 4600 & 6.95 & ?\\
iDPV$^{**}$ & 9.1 & 1.9 & 0.21 & 8.9 & 12900 & 7.24 & 172\\
V393~Sco & 7.8 & 2.0 & 0.25  & 9.4 & 7950 & 7.71 & 253\\
LP~Ara & 9.8 & 3.0 & 0.30 & 15.6 & 9500 & 8.53 & 273\\
V360~Lac  & 7.45 & 1.21 & 0.16 & 9.64$^*$  & 6000 & 10.09 & 322.2\\
AU~Mon & 7.0 & 1.2 & 0.17 & 10.1 & 5750 & 11.11 & 421\\
BR~CMi  & 2.31  & 0.14 & 0.06 & 5.54$^*$ & 4655 & 12.92 & no\\
$\beta$~Lyr & 13.2 & 3.0 & 0.23 & 15.2 & 13200 & 12.94 & 282.4\\
HD~170582 & 9.0 & 1.9 & 0.21 & 15.6 & 8000 & 16.87 & 537 \\
RX~Cas & 5.6 & 1.8  & 0.32 & 23.9$^*$  & 4400 & 32.31 & 516.1\\
V495~Cen & 5.85 & 0.97 & 0.17 & 19.7 & 5000 & 33.49& 1283\\
SX~Cas & 5.1 & 1.5  & 0.29 & 24.42$^*$ & 4000 & 36.56 & ?\\
 \hline
\end{tabular}
\end{center}
\end{table*}%

\begin{figure}[htbp]
\begin{center}
\includegraphics[scale=0.7]{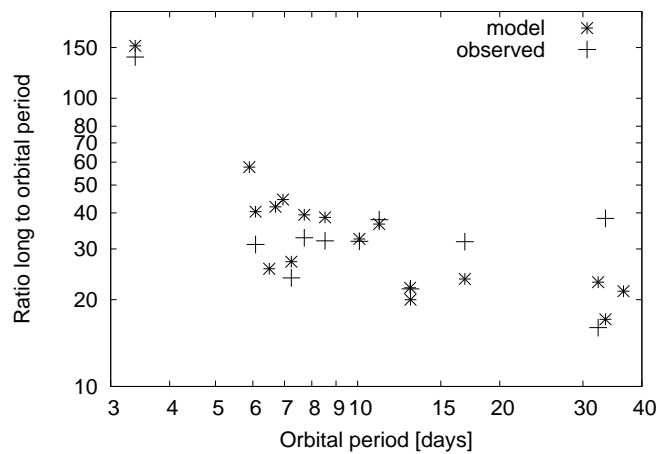}
\caption{Ratio of the long to the orbital period as a function of the orbital period based on the observed data as well as our model. The corresponding data are given in Tables~\ref{DPVsystems} and \ref{DPVcycles2}. {For the system U~Cep, we adopted the lowest observed period in the comparison.}}
\label{compfig}
\end{center}
\end{figure}

\section{Comparing model and observations}\label{comp}
{ In the following, we compare the proposed dynamo model with the available observational data. For this purpose, we present a sample of DPV systems with known physical parameters in subsection~\ref{dpv}, which are compared to our model predictions in subsection~\ref{compare}.
}

\subsection{Systems with known physical parameters}\label{dpv}

{ To test the model outlined above, we need to construct a sample of DPVs with known physical parameters. The largest such sample has been compiled by \citet{Mennickent16}, including masses, radii and effective temperatures both for the donor and the gainer star, in the systems LP~Ara, iDPV  {(OGLE 05155332-6925581)}, HD~170582, V393~Sco, DQ~Vel, AU~Mon and V360~Lac. We also include $\beta$~Lyr in our analysis, even though previously classified as a W~Serpentis system, using the cycle period given by \citet{Harmanec15}. Similarly detailed data were recently obtained for the system V495~Cen \citep{Rosales17}.

From the \citet{Harmanec15} sample, we further include all additional systems with measured cycle periods, i.e. RX~Cas and TT~Hya, as well as the systems SX~Cas, UX~Mon, CX~Dra, V448~Cyg and BR~CMi, where the presence of a cycle is not clear or the cycle period is not yet measured. For these systems, we note that only the stellar masses and orbital periods were directly measured. The radius of the donor star is therefore calculated from the Roche lobel formula (Eq.~\ref{Roche2}). The effective temperature of the donor stars of RX~Cas and SX~Cas is adopted from \citet[][and references therein]{Mennickent16}, the one of TT~Hya based on \citet[][see also similar results by \citet{Etzel88} and \citet{VanHamme93}]{Miller07}, the one of CX~Dra based on \citet[][see also earlier work by \citet{Koubsky80}]{Guinan84}, the one of V448~Cyg based on \citet{Djurasevic09} and the one of BR~CMi was determined by \citet{Harmanec15}. 

{As an additional interesting object, we further include the short-period binary U~Cep, which was initially investigated and analyzed by \citet{Hall75} and more recently for instance by \citet{Manzoori08}. As described by \citet{Manzoori08}, it is a very well known and the presumably most active Algol-type binary, for which they inferred 16 characteristic periods with a Fourier analysis of the data, ranging from $1.41$ up to $73.05$~years. While \citet{Hall75} considered mass loss as the main driver of the variations, \citet{Manzoori08} considers both mass loss and magnetic activity as relevant mechanisms, and also Applegate's model is specifically mentioned in the interpretation. While overall this system is certainly more active than the typical systems studied here, we still include it in our sample to investigate whether some of its periods could be due to magnetic activity.} The pysical parameters of these systems are summarized in Table~\ref{DPVsystems}.
}

\subsection{Comparing observed cycles and predictions}\label{compare}

{ In the following, we apply our dynamo model to the observed systems in the sample presented in Table~\ref{DPVsystems}.  {As already mentioned above, we adopt here $\epsilon_H=1$ and $l_m/H_P=1$ for simplicity. The remaining parameter $\alpha$ is then chosen as $\alpha=0.31$, yielding a rather good agreement of the observational data with the average population. Our value of $\alpha$ lies within the range of possible values given by  \citet{Soon93} and \citet{Baliunas96}, and, similar to the observed scaling relations by \citet{Saar99}, it corresponds to the somewhat lower range of the parameter space.} 

{We however note here that this parameter combination is not unique, and that one could determine appropriate values of $\alpha$ also for other values of $\epsilon_H$ and $l_m/H_P$, while there are certainly parameter combinations that would lead to larger deviations from the observed sample. While we adopt the above choice for definiteness, our main purpose is to show that there is a parameter space for which the model and the data are about consistent. 
}

\footnotetext[1]{While \citet{Harmanec15} still reported the likely presence of a long cycle in UX~Mon, the latter is put in doubt based on recent results by \citet{Mennickent16}, as it cannot be seen in a long baseline. We therefore consider it here as a system without a long cycle.}

The results for the model are given in Table~\ref{DPVcycles2}, and a comparison of the observed and predicted ratios between the long cycle {(as determined from Eq.~\ref{Pfin})} and the orbital period are shown as a function of the orbital period in Fig.~\ref{compfig}. We note that from our sample  of {17} binary stars with well-determined physical parameters, only 10 have a well-measured long cycle. The observational data for the period ratio thus appear to be roughly consistent with a flat relation, with a mean value of 29.7. Model predictions were made for all {17 binaries. If we do not consider the highly extraordinary system U~Cep, the mean value obtained from the models is about $32$, with U~Cep it is about 38. }

Focusing first on the binaries where both model predictions and observational data are available, we find a relatively good agreement. In particular, for the {7 binaries with orbital periods between 5 and 13~days, the maximum deviation between predicted and observed ratio is $30\%$, with the average deviation of order $12\%$. For the 3 binaries with longer orbital periods, the deviation ranges from $26\%$ to about $55\%$.}  Overall, the ratio of long to orbital period appears to slightly decrease with orbital period in the models, while there is no such clear trend in the observational data. We however do note that a clear conclusion cannot be drawn due to the low number statistics. 

{As mentioned above, the system U Cep is a special case, due to its strong activity and the many different periods that were inferred by \citet{Manzoori08}. Our model with the parameters given above would predict an activity cycle of about $1.29$~years, which is comparable to the lowest period inferred by \citet{Manzoori08} of $1.41$~years. If that is the right period to compare with, the difference between observed and predicted cycle is  about $8\%$. However, the comparison clearly needs to be treated with caution, due to the many cycles found in the system and the generally strong activity, which will certainly involve other mechanisms as well.}

\begin{table*}[!htbp]
\begin{center}
\caption{Comparison of model predictions with observational results assuming $\epsilon_H=1$, $l_m/H_P=1$ and $\alpha=0.31$. The table includes the observed and predicted ratios of cycle to orbital period, the ratio between the predicted and observed cycle, the maximum orbital period variation due to magnetic activity, the relative maximum change in the donor radius due to magnetic activity as well as the maximum modulation of the mass transfer rate due to magnetic activity. We note that, when available, we used the observed cycle period in the calculation of the mass transfer rate, while otherwise employing the model prediction. These cases are marked below with an asterix. }
\begin{tabular}{c|c|c|c|c|c|c}\label{DPVcycles2}
binary & $\left(P_{\rm cycle}/P_{\rm orb}\right)_{\rm obs}$ & $\left(P_{\rm cycle}/P_{\rm orb}\right)_{\rm model}$  & $P_{\rm cycle, model}/P_{\rm cycle, obs}$ & $\left(\Delta P/P_{\rm orb}\right)_{\rm max}$ & $\left(\Delta R/R_2\right)_{\rm max}$ & $\dot{M}_{\rm max}$  [$M_\odot$~yr$^{-1}$] \\ \hline
U~Cep & 152-7889 & $139$ & max. 0.92 & $1.2\times10^{-7}$ & $6.7\times10^{-8}$ &  $8.4\times10^{-10}$   \\
UX~Mon & - & 57.7 & - & $1.7\times10^{-6}$ & $1.4\times10^{-6}$ & $5.7\times10^{-9}$\\
DQ~Vel  & 31.1 & 40.4 & 1.3 & $3.0\times10^{-6}$ & $4.2\times10^{-6}$ & $5.1\times10^{-8}{ }^*$\\
V448~Cyg & - & 25.6 & - & $3.0\times10^{-5}$ & $3.1\times10^{-5}$ & $1.7\times10^{-6}$\\ 
CX~Dra & - & 42.0 & - & $3.3\times10^{-5}$ & $2.0\times10^{-5}$ & $1.9\times10^{-7}$\\
TT~Hya & - & 44.5 & - & $4.2\times10^{-7}$ & $1.3\times10^{-6}$ & $2.6\times10^{-9}$\\
iDPV & 23.8 & 27.1 & 1.1 & $1.3\times10^{-5}$ & $2.1\times10^{-5}$ & $4.4\times10^{-7}{ }^*$ \\
V393~Sco & 32.8 & 39.4 & 1.2 & $3.1\times10^{-6}$ & $4.8\times10^{-6}$ & $4.3\times10^{-8}{ }^*$\\
LP~Ara & 32 & 38.6 & 1.2 & $1.7\times10^{-5}$ & $1.3\times10^{-5}$ & $2.1\times10^{-7}{ }^*$\\
V360~Lac  & 31.9 & 32.5 & 1.0 & $6.4\times10^{-6}$ & $1.2\times10^{-5}$ & $7.1\times10^{-8}{ }^*$\\
AU~Mon & 37.9 & 36.6 & 1.0 & $4.3\times10^{-6}$ & $7.8\times10^{-6}$ & $3.4\times10^{-8}{ }^*$\\
BR~CMi & - & 22.0 & - & $1.6\times10^{-5}$ & $5.6\times10^{-5}$ & $1.9\times10^{-7}$\\
$\beta$~Lyr & 21.8 & 20.0 & 0.91 & $1.1\times10^{-4}$ & $8.5\times10^{-5}$ & $1.9\times10^{-6}{ }^*$\\
HD~170582 & 31.8 & 23.6 & 0.74 & $3.3\times10^{-5}$ & $5.6\times10^{-5}$ & $4.0\times10^{-7}{ }^*$\\
RX~Cas & 16 & 23.0 & 1.4 & $2.4\times10^{-5}$ & $3.3\times10^{-5}$ & $2.0\times10^{-7}{ }^*$\\
V495~Cen & 38.3 & 17.1 & 0.45 & $6.6\times10^{-4}$ & $3.2\times10^{-4}$ & $7.8\times10^{-7}{ }^*$\\
SX~Cas & - & 21.4 & - & $4.5\times10^{-5}$ & $6.3\times10^{-5}$ & $2.3\times10^{-7}$\\
 \hline
\end{tabular}
\end{center}
\end{table*}%

{For the verification of our model, we further} compare its predictions to the numerical simulation by \citet{Blackman01}, to our knowledge the only numerical simulation so far where the non-linear dynamo phase has been modeled for an AGB star. While their model includes strong differential rotation, we expect here a different regime of rather rigidly rotating stars as a result of strong tidal interaction. Nevertheless, a comparison may be instructive to see if the order of magnitude is about right. They consider an AGB star with $3$~M$_\odot$, a radius of $4.3$~R$_\odot$, a luminosity of about $20$~L$_\odot$ and a rotation period of about $14$~days around the convection zone. Inserting these numbers in our model, we obtain an expected ratio of about 33.3, while the ratio found in their simulations corresponds to a value of $10.4$. From this comparison, we may thus conclude that the accuracy of our model lies within about a factor of $3$. {Still, we note that such a comparison has to be treated with caution, due to uncertainties present even in the modeling of the solar dynamo.} In the limits of such uncertainties, we may cautiously conclude that a dynamo mechanism can potentially produce long periods similar to the observed ones, though of course further investigation is necessary. 

In Table~\ref{DPVcycles2}, we further provide predictions based on Applegate's model for the maximum ratios $\Delta P/P_{\rm orb}$ due to magnetic activity, with typical values of $10^{-5}-10^{-6}$. The highest possible value is found in the system  V495~Cen, with a maximum ratio of $6.6\times10^{-4}$. We note that due to the low magnitude of these variations and as these are upper limits, a direct comparison with observations is however difficult, as we cannot separate the phenomenon from other effects like binary evolution, which also produces period variations and a time-dependence of the mass transfer rate. Similarly, the maximum variation in the relative stellar radius due to magnetic activity is found to be of the order of $10^{-5}-10^{-6}$, again with a maximum value of $3.2\times10^{-4}$ in case of V495~Cen. The predicted modulations of the mass transfer rate range from a maximum of $1.9\times10^{-6}$~M$_\odot$~yr$^{-1}$ ($\beta$~Lyr) to a minimum value of $2.6\times10^{-9}$~M$_\odot$~yr$^{-1}$ (TT~Hya). 

{For the highly active system U~Cep, it is interesting to note that our model predicts a rather weak influence of the Applegate mechanism, with period variations $\Delta P/P_{\rm orb}$ of the order $10^{-7}$ and a modulated accretion rate of the order $10^{-9}$~M$_\odot$~yr$^{-1}$. The latter provides another indication that Applegate's model is probably not the main reason for the activity in that system.} As mentioned above, these values correspond to an average over the dynamo cycle, and the actual variations at a given time may be higher than the average. It is to be noted, though, that the value of the average corresponds to an upper limit. 

As discussed in the previous section, the presence of a long cycle is not always fully clear. As reported by \citet{Harmanec15}, the systems V448~Cyg and BR~CMi show no signs of such a cycle at this point. We note that these systems appear as somewhat unusual. In particular, the donor in V448~Cyg has a very high mass of 13.7~M$_\odot$, while the donor in BR~CMi has a very low mass of $0.14$~M$_\odot$. These systems thus correspond to the extreme cases for the donor mass range considered here, and it is at least conceivable that in this physical regime, the dynamo process is different or other stellar processes are more important. In case of BR~CMi, an additional concern is the very low mass ratio of about 0.06, implying a large Lubow-Shu critical radius. While most DPVs were previously found to be tangential-impact systems \citep{Mennickent16}, this system would be more likely to form a disk, implying a different energy dissipation mechanism. The latter is potentially more gradual, and may thus wash out some of the originally present fluctuations due to the dynamo. The recent investigation by \citet{Mennickent16} {found that a constant orbital period reproduces well the light curve of UX\,Mon during 57 years, casting doubts about the previously reported period changes for this star. Moreover, they found no evidence for a long cycle.} We note that all three of these systems have rather extreme mass ratios, which can potentially be relevant for the interpretation.

For the systems TT~Hya and SX~Cas, the presence of a long cycle is unclear, but not strongly ruled out. Our model predicts here typical ratios of the long to orbital period of 41.5 and 17.1, respectively. The expected modulation of the accretion rate in TT~Hya is quite on the low side, with $2.6\times10^{-9}$~M$_\odot$~yr$^{-1}$, while one expects about $2.3\times10^{-7}$~M$_\odot$~yr$^{-1}$ for SX~Cas. These relatively low values may partly contribute to the difficulty to detect a long cycle, though we encourage additional observations in particular for SX~Cas to either confirm a long cycle or to provide stronger constraints. Finally, we note that for the system CX~Dra, indications for a long cycle are present, but its duration is not well constrained. A better determination of the cycles will thus be valueable for comparison with the model predictions given here.

}

\section{Discussion and conclusions}\label{discussion}
In this manuscript, we have considered the magnetic activity as a potential explanation for the long period in DPVs, {which exhibit cyclic light variations on a timescale an order of magnitude larger than the corresponding orbital time scale, many of them with a characteristic ratio of about $3.5\times10^1$ \citep{Mennickent03}.} As a result of efficient synchronization via dynamical tides, we expect rapid rotation of the donor star, with its rotation period equal to the period of the binary system \citep{Tassoul87, Zahn89a, Zahn89b}. The resulting rapid rotation, corresponding to $10-30\%$ of the Keplerian value, provides ideal conditions to drive a magnetic dynamo. 

{ To estimate the dynamo cycle within the star, we have adopted the relation proposed by \citet{Soon93} and \citet{Baliunas96} between the activity cycle and the orbital period, and estimated the dynamo number in the stellar interior. Based on Applegate's mechanism, particularly the formulation provided by \citet{Volschow16}, we further estimated the impact of the magnetic dynamo on the stellar interior, including changes in the quadrupole moment, the stellar radius and the resulting modulation of the mass transfer rate of the star. {This mechanism is known to produce orbital period variations in magnetically active binaries, and we suggest here that it can also lead to a time-variable mass transfer rate in semi-detached systems, such as the DPVs. We have shown that the expected orbital period variations are negligible for the DPVs, consistent with the observational results, thus leaving the variable mass transfer as the main relevant observable.}

This model has been applied to a sample of close binaries with at least one massive star and known physical parameters, including the DPV candidates outlined by \citet{Mennickent16} and additional systems described by \citet{Harmanec15}. In particular for the systems with rotation periods {between 5 and 13~days, we find a good agreement with our model, with average deviations of the long cycle of the order $12\%$, while larger variations up to $55\%$ are possible for longer rotation periods. This match is obtained using the same basic assumptions for the whole set of targets and with only one adjustable parameter, $\alpha$.} As an additional estimate of our model uncertainties, we have compared it to the numerical simulation of an AGB dynamo by \citet{Blackman01}, to date unfortunately the only numerical simulation investigating the non-linear phase of the dynamo in this regime, finding an uncertainty of about a factor of $3$. {Despite the uncertainties both in the model as well as within the dynamo simulations}, it shows that the required orders of magnitudes can be produced via a dynamo mechanism.

Particularly interesting are the objects where no long cycle has been found to date, including V448~Cyg and BR~CMi \citep{Harmanec15}. These objects are at the extreme ends for the masses of the donor stars, with $13.7$~M$_\odot$ for V448~Cyg and $0.14$~M$_\odot$ for BR~CMi. In case of BR~CMi, we further note that the low mass ratio of 0.06 implies a large Lubow-Shu critical radius, potentially leading to the formation of an extended disk around the primary star. In case of TT~Hya and SX~Cas, the presence of a long cycle is also unclear, but still feasible. In case of UX~Mon, the analysis by \citet{Mennickent16} is also not compatible with a long cycle. We note that all three of these systems have rather extreme mass ratios, which may be relevant for the interpretation. The system CX~Dra shows signs of a long cycle, but the duration is not well constrained, therefore limiting the possibility of a model comparison.} We also note that in a few cases, the long cycle has turned out to be non-periodic. The most extreme case is LMC~SC6~57364. \citet{Mennickent05} found that between JD~2448800 and 2450000, the long term period was $340$~days, while it shortened to $270$~days around JD~2450500, implying a variability amplitude of $20\%$. \citet{Poleski10} report for this object $dP_2/dt=-0.01724\pm0.00040$. Such variability is also known from the magnetically active RS~CVn stars \citep{Lindborg13}, and activity variations are known also from solar analogues \citep{Kapyla16}, and are thus generally consistent with the idea of the long cycle being driven by a dynamo mechanism.

{
In general, we should note that the current number of objects in our sample with known physical parameters and known long cycles is only {11}, thus limiting the ability to identify clear trends in the observational sample and limiting the comparison with model predictions. In addition, even within these systems, the physical parameters are not always well constrained, and further work on their characterization is certainly needed. Even in well-studied systems like AU~Mon with a very good eclipsing light curve from CoRoT and very systematic ground-based V photometry \citep{Lorenzi80, Desmet10}, minor uncertainties are present in the physical parameters. For instance, \citet{Desmet10} have solved the light curve assuming a semi-detached configuration and interpreting the residuals from the fit as due to short-periodic oscillations, finding a radius of $5.6\pm0.5$~R$_\odot$ for the more massive star. \citet{Djurasevic10} solved the same light curve assuming also a flat accretion disk including 3 spots, yielding a radius of $5.1\pm0.5$~R$_\odot$, thus roughly consistent within the errors. \citet{Mimica12} instead assumed a clumpy and asymmetric accretion disk, and \citet{Atwood12} considered additional contributions from the gas stream. All of these studies are however based on the mass ratio of $0.17\pm0.03$ estimated by \citet{Desmet10}, which limits the accuracy that can be reached at this point. {In other few systems, the mass of the B-type star is inferred from the spectral type assuming a main sequence classification \citep{Mennickent16}.}

{We also investigated the short period binary U~Cep \citep{Hall75, Manzoori08}, as it has well-determined physical parameters and is the perhaps most active Algol-type system known to date. We found that our model can potentially explain the shortest period inferred in that system, even though the predicted magnitude of the period variation as well as the modulation of the accretion rate are rather weak. It is thus conceivable that the activity in the system is mostly driven by mass loss, as originally proposed by \citet{Hall75}.}

In this paper we have shown that the DPV long cycles could be caused by changes in the donor 
radius driven by the Applegate mechanism; consistently they should produce cyclic mass transfer variations 
potentially observable by changes in the accretion luminosity. We notice however that due to the presence of the disc,
the gas stream cannot hit directly the star, but hits the disc at its outer edge, producing a hotspot or hotline,
which is revealed in light curve models \citep{Mennickent13}  and hydrodynamical simulations 
\citep{Bisikalo98, Bisikalo99, Bisikalo03}. It is then possible that an increase 
of the energy released at the hotspot produces a stronger hotspot wind at long cycle maximum, as modeled by \citet{Rensbergen08}.  This strengthened  wind can be seen as the extra light emitting source reported for instance in V393 Scorpii \citep{Mennickent12}, DQ Velorum \citep{Barria14} and $\beta$ Lyrae \citep{Harmanec96, Hoffman98, Ak07}. In this case the hotspot should be brighter at long cycle maximum, something revealed by Doppler tomography in the DPV HD 170582 \citep{Mennickent16b}. More work is needed in this line to firmly validate our proposed scenario.

The possibility to drive dynamos in Algol-type systems is nothing new, but was already proposed by \citet{Sarna97} and \citet{Soker02}. Resulting variations of the mass transfer rate have been suggested by \citet{Bolton89} and \citet{Meintjes04}, considering various mechanisms, including the direct impact of the magnetic field on the stellar surface and the stream as well as the potential presence of cool spots due to rising magnetic bubbles, which may interact with the mass stream and change both the densities and temperatures close to L1. We here consider the potential contribution resulting from the Applegate mechanism \citep{Applegate87, Volschow16}, implying cyclic variations of the quadrupole moment, the stellar radius and the mass transfer rate, finding typical values of the modulation of $10^{-7}-10^{-8}$~M$_\odot$~yr$^{-1}$.
}

Our results have been derived based on simplifying assumptions, considering mean properties of the stellar interior and the approximation of mixing length theory. Also the use of the Roche lobe formula by \citet{Paczynski71} is a simplifying assumption, as the donor star may flatten as a result of rapid rotation, potentially altering the equipotential surfaces. Within the limit of these uncertainties, it appears at least plausible that a dynamo cycle with properties similar to the observed cycle will emerge, and independent of the details, we certainly expect magnetic activity in these type of systems. A piece of evidence pointing in this direction is the detection of chromospheric activity in the donor of the DPV V393 Scorpii \citep{Mennickent12}. { For a better understanding of dynamos in these type of systems, it is however necessary to first better understand dynamos in giant and AGB stars. While there is some theoretical work pursued by \citet{Sarna97} and \citet{Soker02}, the only simulation exploring the non-linear dynamo in AGB stars is based on \citet{Blackman01}, while \citet{Dorch04} investigated the linear regime including the growth of the magnetic field. It will however be desirable to further explore the dependence on various physical parameters, including the amount of rotation and different profiles of the rotational velocity, as well as the masses and structure of the stars. On the long term, one should pursue a joint modeling of the dynamo in the donor star along with the whole binary system (the primary potentially approximated via a point particle), to obtain a better understanding of the dynamical evolution in such situations.
}

The model proposed here can be tested probing the magnetic activity of the donor star using polarimetry. In addition, one may search for indirect confirmations or constraints. As an example, \citet{Mennickent16} have discussed 3 RS~CVn systems that were previously misidentified as DPVs. While these systems are detached and not in the mode of accretion, a long cycle with very similar properties as in the case of DPVs has been inferred from possible variable surface spots, providing an indirect confirmation of the proposed scenario. An identification of more such systems would thus be desirable in order to probe whether it is indeed a universal phenomenon. Time-dependent measurements of the accretion rates will help to confirm a potential modulation of the accretion rate via dynamo cycles and link the variation more closely to the properties of the donor star. { In addition, the extension of such samples with known, well-constrained physical properties of the binaries and known, well-understood long cycles will be crucial to build up better statistics and to more strongly constrain the physical picture driving the long period.}

\begin{acknowledgements}
{We thank the two referees of our paper for insightful comments that helped to improve our manuscript.} DRGS thanks for funding through Fondecyt regular (project code 1161247) and through the ''Concurso Proyectos Internacionales de Investigaci\'on, Convocatoria 2015'' (project code PII20150171). R.E.M. acknowledges support by VRID-Enlace 214.016.002-1.0 and the BASAL Centro de Astrof\'isica y Tecnolog\'ias Afines (CATA) PFB-06/2007. 

\end{acknowledgements}


\end{document}